\documentclass[12pt]{article}

\usepackage{graphicx}

\usepackage{hyperref}
\hypersetup{colorlinks=true, citecolor=blue, linkcolor=blue, urlcolor=blue}

\usepackage{authblk}

\usepackage{subcaption}

\usepackage[font=scriptsize, labelfont=bf]{caption}

\usepackage{sectsty}
\sectionfont{\normalsize}
\subsectionfont{\small}
\subsubsectionfont{\small}

\usepackage{cite}

\providecommand{\keywords}{\small\textbf{Keywords: }}

\title{\large \textbf{Production and spectroscopic investigation of Mercury and Radon isotopes produced in complete fusion reaction and multi-nucleon transfer reaction at MASHA facility}}

\author[1,a)]{R. Pandey}
\affil[1]{Graduate Engineer Trainee, Larsen $\&$ Toubro Limited, Faridabad, Haryana, India.}
\affil[a)]{Corresponding author: rishav160999@gmail.com}

\date{}

\begin{document}

\maketitle

\begin{abstract}
In this paper, the production and spectroscopic investigation of Mercury and Radon isotopes was performed using complete fusion reactions neutron evaporation residues and multi-nucleon transfer reaction at the mass-separator MASHA. The MASHA setup is installed on the beam line of Cyclotron U-400M at Flerov Laboratory of Nuclear Reactions (FLNR) in Joint Institute for Nuclear Research (JINR), Dubna, Russia. The isotopes produced in complete fusion reactions $^{148}Sm(^{40}Ar,xn)^{188-x}Hg$, $^{166}Er(^{40}Ar,xn)^{206-x}Rn$ and multi-nucleon transfer reaction $^{48}Ca$ + $^{242}Pu$ were passed through the magneto-optical system of MASHA setup with charge state Q=+1 and were separated on the basis of their mass to charge ratio. For the detection of these isotopes, a position sensitive Si detector was used. Further, the experimental data obtained were analysed and spectroscopic investigations were carried out.
\end{abstract}

\keywords{Spectroscopic investigation, Complete Fusion Reaction, Multi-Nucleon Transfer Reaction, Mass separator, MASHA, Position sensitive detector.}

\section{INTRODUCTION}
The MASHA (Mass Analyzer of Super Heavy Atoms) setup has been designed as a mass-separator with the resolving power of about 1700, which allows mass identification of super-heavy nuclides. The setup uses the solid ISOL (Isotope Separation On-Line) method. The work on the project means the analysis of real data collected from the experiments of complete fusion reactions neutron evaporation residues $^{40}Ar$ + $^{148}Sm$ $\rightarrow$ $^{188-x}Hg$ + $xn$, $^{40}Ar$ + $^{166}Er$ $\rightarrow$ $^{206-x}Rn$ + $xn$  and multi-nucleon transfer reaction $^{48}Ca$ + $^{242}Pu$ using the $\alpha$-decay chains from the position sensitive Si detector.

An analysis was done on the experimental data from the above stated reactions to calculate the masses of identified isotopes(which are results of the above nuclear reactions), their half life, Alpha Branching Ratio (ABR), energy of $\alpha$-decay $(E_{alpha})$ and their probability to decay with a specific amount of energy. Using these experimental data, one dimensional $\alpha$-decay energy spectrum was plotted and its peak analysis was performed. Further, using the 1D histograms ($\alpha$-decay energy spectrum), a heatmap (two dimensional energy-position graph) was obtained for the isotopes. Finally, a comparison was drawn between the theoretical values and experimental values of $E_{alpha}$, and conclusions were made. The former was obtained from the nuclide chart, while the latter was obtained by performing experiments at the MASHA facility.

\section{LAYOUT OF MASHA SETUP AND ITS MAIN PARTS}
The MASHA setup is shown in Fig.\ref{Schematic of the MASHA facility: D1, D2, D3a, D3b are dipole magnets, Q1, Q2, Q3 are quadrupole lenses, S1, S2 are sextupole lenses.}. It consists of a Target Box with a Hot Catcher, an ECR Ion Source, Magneto-Optical System (or mass to charge ratio analyzer), and a Position Sensitive Si Detector (Detection and Control System). It also consists of an intermediate plane (F1), and for detailed monitoring of isotopes an additional strip detector has been installed at this middle plane also, apart from the detector present at focal plane (F2). Each part of the MASHA facility is explained in detail below. For in-depth knowledge on MASHA setup, one can refer this paper \cite{vedeneev2017current}.

\begin{figure}[h]
\centering
\includegraphics[scale=0.6]{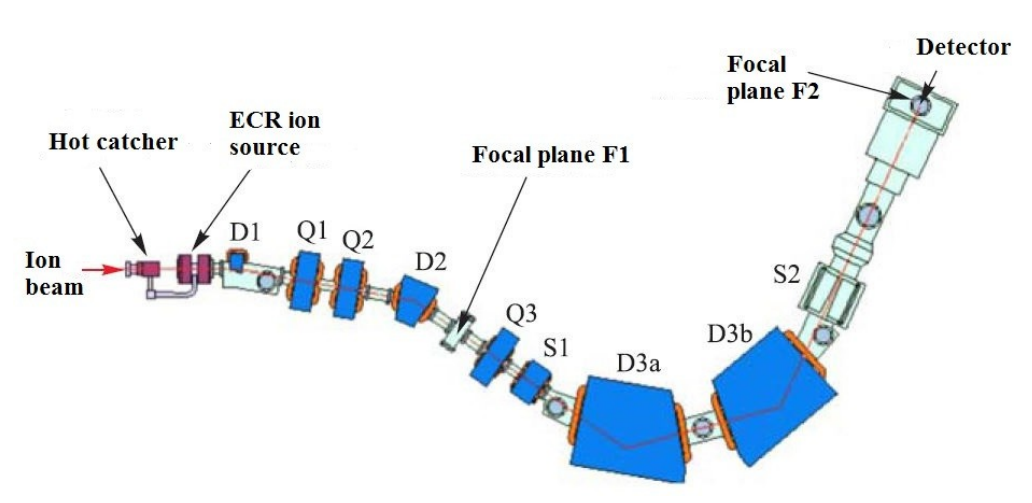}
\caption{Schematic of the MASHA facility: D1, D2, D3a, D3b are dipole magnets, Q1, Q2, Q3 are quadrupole lenses, S1, S2 are sextupole lenses.}
\label{Schematic of the MASHA facility: D1, D2, D3a, D3b are dipole magnets, Q1, Q2, Q3 are quadrupole lenses, S1, S2 are sextupole lenses.}
\end{figure}

\subsection{TARGET BOX}
The target box consists of a rotating disc divided into 6 sectors, which are sputtered with target material(s) as shown in Fig.\ref{Rotating target disc inside the target box.}. The disc rotates with a frequency of 25Hz \cite{vedeneev2017current}. The high energetic projectile particle ejected from U-400M cyclotron collides with the target material present in rotating disc to induce some kind of nuclear reaction. The products of the nuclear reaction are stopped by the hot catcher which is discussed below.

\subsection{HOT CATCHER}
\begin{figure}[h]
\centering
\begin{subfigure}[h]{0.49\textwidth}
\centering
\includegraphics[scale=0.535]{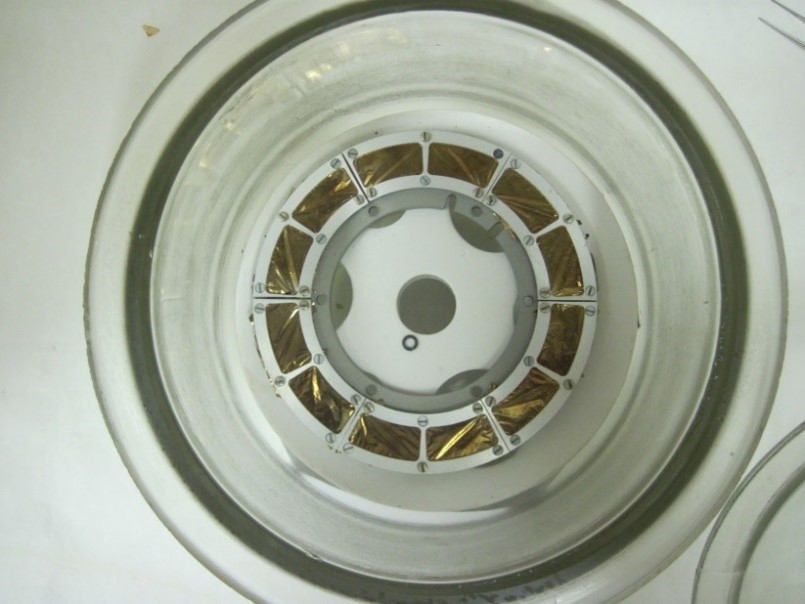}
\caption{Rotating target disc inside the target box.}
\label{Rotating target disc inside the target box.}
\end{subfigure}
\hfill
\begin{subfigure}[h]{0.49\textwidth}
\centering
\includegraphics[scale=0.535]{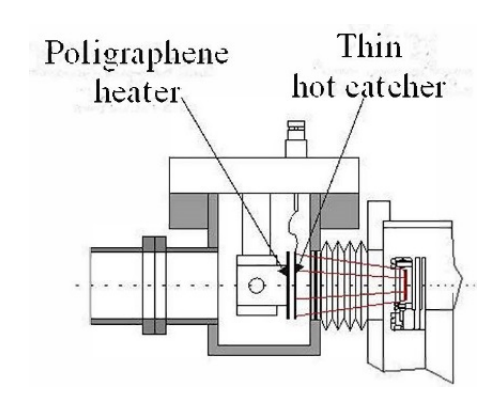}
\caption{A thin hot catcher installed just before poly-graphene heater.}
\label{A thin hot catcher installed just before poly-graphene heater.}
\end{subfigure}
\caption{}
\end{figure}

\begin{figure}[h]
\centering
\includegraphics[scale=0.7]{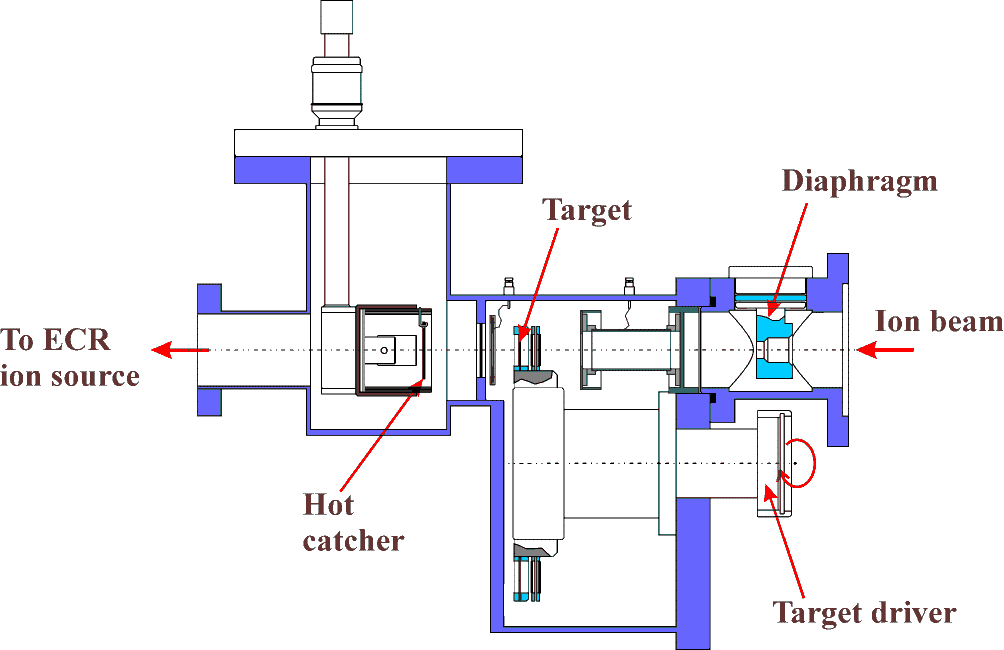}
\caption{Complete schematic of target box with hot catcher.}
\label{Complete schematic of target box with hot catcher.}
\end{figure}

The diagram of hot catcher is shown in Fig.\ref{A thin hot catcher installed just before poly-graphene heater.}. The below setup mainly consists two components, one is poly-graphene heater and the other one is absorber material. The latter is usually made up of thin film of graphite or carbon nanotubes heated by the former upto a temperature of $1800-2000^oC$. As depicted in figure the absorber is installed in front of the heater at a distance of 2 mm along the beam axis of MASHA \cite{mamatova2019study}. The products of the nuclear reaction is stopped by the absorber material, vaporized to gaseous form and are passed to the ECR ion source. Fig.\ref{Complete schematic of target box with hot catcher.} shows the complete schematic of target box with hot catcher.

\subsection{ECR ION SOURCE}
The ECR (Electron Cyclotron Resonance) ion source with a microwave oscillation frequency of 2.45 GHz \cite{vedeneev2017current, mamatova2019study, rodin2014separation, rodin2014masha}, acts as an ionization chamber of MASHA spectrometer. It ionizes the atoms of gaseous isotopic products of nuclear reaction to a charge state Q=+1, and accelerates them to an energy of 38 KeV using three electrode system \cite{vedeneev2017current}. The ionized atoms gets converted to beam and are then separated by magneto-optical system of the MASHA spectrometer.

\subsection{MAGNETO-OPTICAL SYSTEM}
The magneto-optical system separates the beam of ions on the basis of their mass to charge ratio. The magnetic separation of heavy nuclei is performed using four dipole magnets(D1, D2, D3a, D3b), three quadrupole lenses (Q1, Q2, Q3) and two sextupole lenses (S1, S2) as shown in Fig.\ref{Schematic of the MASHA facility: D1, D2, D3a, D3b are dipole magnets, Q1, Q2, Q3 are quadrupole lenses, S1, S2 are sextupole lenses.} \cite{mamatova2019study}. Once, the heavy nuclei gets separated they are then detected at different strips of position sensitive Si detector.

\subsection{POSITION SENSITIVE Si DETECTOR}
The position sensitive Si detector is a multiple detector system used to detect the separated heavy nuclei. It is installed at the focal plane (F2) of the MASHA setup. A clear view of the position sensitive Si detector is shown in Fig.\ref{Position sensitive Si detector. 1-front detector, 2-upper detector, 3-lower detector, 4-lateral detector.}. The front detector has a dimension of 240x35 mm$^2$ and it consists of 192 strips. The upper and lower detector consists of 64 strips each while the left and right lateral detector consists of 16 strips each. Each strip has a width of 1.25 mm and each detector has a thickness of 0.3 mm \cite{rodin2014separation, rodin2014masha, rodin2020features}.

\begin{figure}[h]
\centering
\includegraphics[scale=.372]{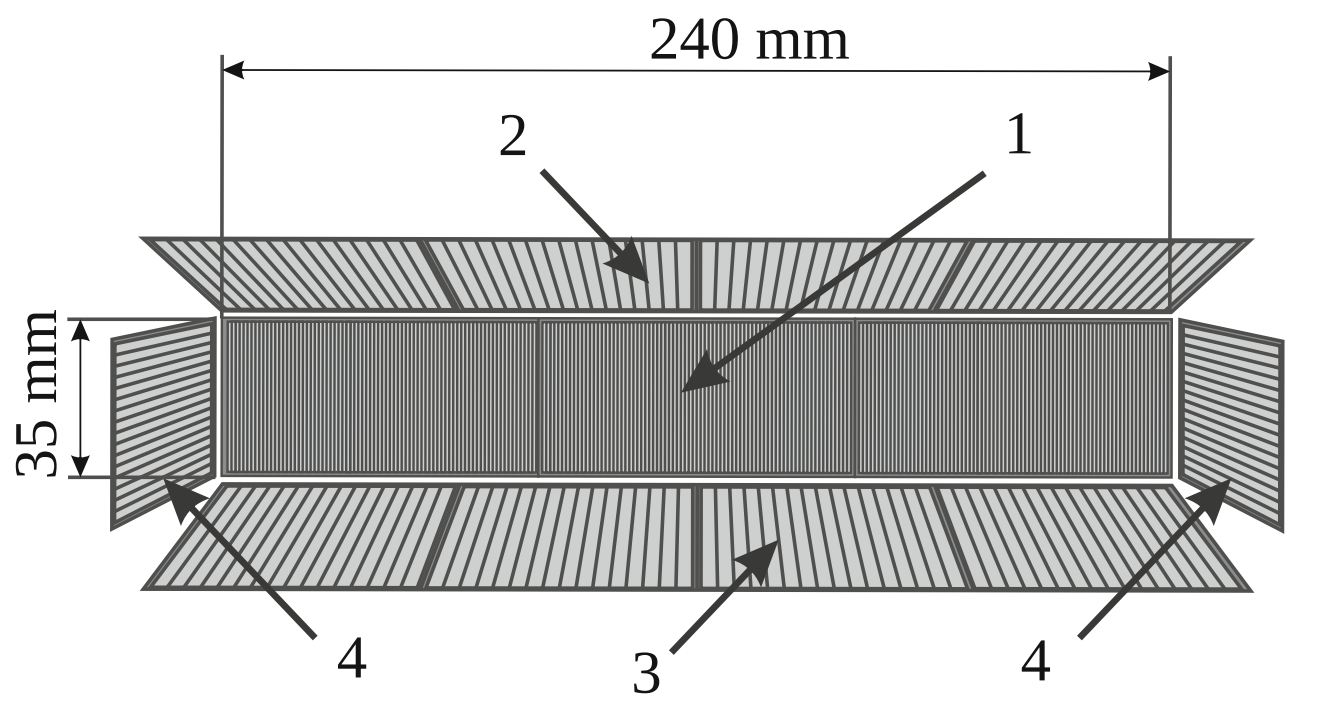}
\caption{Position sensitive Si detector. 1-front detector, 2-upper detector, 3-lower detector, 4-lateral detector.}
\label{Position sensitive Si detector. 1-front detector, 2-upper detector, 3-lower detector, 4-lateral detector.}
\end{figure}

\section{SCIENCE BEHIND THE EXPERIMENT}
The test experiments carried out at FLNR are:
\begin{enumerate}
\item $^{40}Ar$ + $^{148}Sm$ $\rightarrow$ $^{188-x}Hg$ + $xn$
\item $^{40}Ar$ + $^{166}Er$ $\rightarrow$ $^{206-x}Rn$ + $xn$
\item $^{48}Ca$ + $^{242}Pu$ $\rightarrow$ (Any element whose Z varies from 20-114)
\end{enumerate}
The first and second nuclear reactions are complete fusion reactions neutron evaporation residues. In such types of reactions the product nucleus formed has no. of protons exactly equal to no. of protons of projectile particle + no. of protons of target nucleus. Here, all the nucleons participate in the reaction. While the third one is a Multi-Nucleon Transfer Reaction (MNTR). In such nuclear reactions different nuclides can be formed whose atomic no. ranges from atomic no. of projectile particle to sum of atomic nos. of projectile particle + target nucleus. This means not all nucleons participate in this reaction and can lead to formation of any possible product. The N/Z ratio of product nucleus can be higher or lesser than than the optimal ratio required for its stability (i.e. It can be proton rich or neutron rich).

The U-400M cyclotron installed at FLNR, JINR is used to accelerate projectile particle ($^{40}Ar$ \& $^{48}Ca$) to a very high velocity, with an energy \~ 240 MeV (for $^{40}Ar$ + $^{148}Sm$) and with energy \~ 198 MeV ($^{40}Ar$ + $^{166}Er$). The high energetic projectile particle enters into the MASHA setup and induce a nuclear reaction by colliding with target material sputtered in rotating disc present in target box of MASHA facility. The products of nuclear reaction are isotopes of Hg (for $^{40}Ar$ + $^{148}Sm$)  and Rn (for $^{40}Ar$ + $^{166}Er$ and $^{48}Ca$ + $^{242}Pu$) which are stopped by the absorber material of hot catcher.

The absorber material is generally made up of thin film of graphite or carbon nanotubes which is heated to around $1800-2000^oC$ by means of IR radiations coming out from poly-graphene heater as well as by a direct current passing through the absorber. This absorber stops the isotopic products of nuclear reaction, vaporizes them and their respective atoms diffuses through this absorber material into the vacuum volume of the hot catcher. Moving along the vacuum pipe, they reach the ECR ion source \cite{rodin2014separation}. This ECR ion source acts as an ionization chamber of MASHA setup where the atoms of gaseous isotopic products gets ionized to charge state Q=+1 and further they are accelerated with the help of three electrode system. (The three electrode system consists of one positive electrode, one negative electrode and one more negative electrode. Hence, an electric field is established from positive electrode to negative electrode. So, when a charged particle (here, ion) moves in the direction of electric field, it gets accelerated.

The product isotopes are then separated by their M/Q ratio in the magneto-optical system of MASHA setup and at last they reach to the focal plane (F2) of the position sensitive Si detector and are detected at different strip numbers. (i.e. different isotopes are detected at different strip numbers).

Now, the science is that the separated heavy nuclei undergoes
$\alpha$-decay to produce daughter nuclei and it's exactly the alpha particles (with different energies) given out by both parent nucleus and its daughter nuclei which are detected at unique strip nos. of position sensitive Si detector. The detector used is a hybrid pixel detector of the TIMEPIX type, with high resolution and sensitivity which can detect even a single $\alpha$ or $\beta$ particle. So, from the experimental data, we plot $\alpha$-decay energy spectrum for those strips where an isotope was detected. From this spectrum ($\alpha$-decay energy vs. No. of counts) we analyse the prominent peaks and calculate their $\alpha$-decay energy ($E_a$) values. The base peak with maximum no. of $\alpha$ particles (with constant energy) is our point of interest as it could be any one of the separated nuclei. Now, using the table of nuclides, we find which isotope (of product of nuclear reaction) undergoes $\alpha$-decay with energy very close to it. That particular isotope will be the one detected at a unique strip number. Ones, the isotope gets detected, then its mass, ABR, daughter nuclei can easily be investigated using the table of nuclides. In the same way, one can detect all the isotopes of an element which is the product of a nuclear reaction.

In this work, a two dimensional energy-position graph (called heatmap) for all three test experiments has also been analysed. This graph gives a clear understanding that which isotope is detected at which strip no. and corresponding to that particular isotope, how many alpha particles (counts) are detected with a constant energy. This constant energy is the energy of $\alpha$-decay of that isotope.

\section{SPECTROSCOPIC INVESTIGATION OF MERCURY ISOTOPES USING FULL-FUSION REACTION $^{148}Sm(^{40}Ar,xn)^{188-x}Hg$}
The complete fusion reaction of $^{148}Sm(^{40}Ar,xn)^{188-x}Hg$ was carried out at MASHA setup. The target material sputtered in rotating disc was $^{148}Sm$ and the products of the nuclear reaction were isotopes of Hg. However, only the long-lived isotopes of Hg were detected whose half-life was greater than average separation time (1.8±0.3 s) used by ISOL method for this reaction \cite{vedeneev2017current, rodin2014separation}.

\subsubsection{PRODUCTION OF $^{180}Hg$}
\begin{figure}[h]
\centering
\includegraphics[scale=0.5]{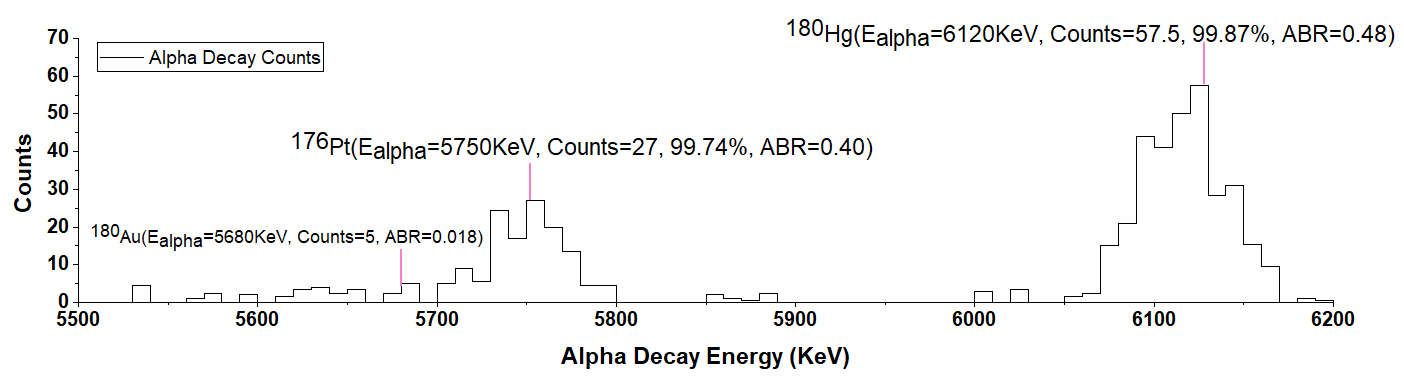}
\caption{$\alpha$-decay energy spectrum of $^{180}Hg$ and its decay products.}
\label{alpha-decay energy spectrum of 180Hg and its decay products.}
\end{figure}

In the $\alpha$-decay energy spectrum of $^{180}Hg$, the most prominent peaks corresponding to parent nucleus and daughter nuclei were analysed. Here, the label $^{180}Hg(E_{alpha}=6120KeV, Counts=57.5, 99.87\%, ABR=0.48)$ gives a lot of information. `$E_{alpha}=6120KeV$' is the energy of $\alpha$-particle released during the alpha decay of isotope $^{180}Hg$. $Counts=57.5$ indicates the no. of $\alpha$-particles detected at a strip of detector. $ABR=0.48$ is the Alpha Branching Ratio of $^{180}Hg$ which means `The probability of $^{180}Hg$ to undergo $\alpha$-decay is 0.48'. It is to be noted that a nucleus can decay with multiple $\alpha$-decay energies and for same reason, here $99.87\%$ means 99.87\% of $^{180}Hg$ decays with energy $=$ 6120 KeV. Now, coming to the analysis part, in Fig.\ref{alpha-decay energy spectrum of 180Hg and its decay products.} we observe that $^{180}Hg$($t_\frac{1}{2}=2.58s$) is peaked at 6120 KeV and its daughter nuclei $^{176}Pt$($t_\frac{1}{2}=6.3s$), formed due to $\alpha$-decay of $^{180}Hg$ and $^{180}Au$($t_\frac{1}{2}=8.1s$), formed due to Electron Capture (EC) in $^{180}Hg$ are peaked at 5750 KeV and 5680 KeV respectively.

\subsubsection{PRODUCTION OF $^{181}Hg$}
\begin{figure}[h]
\centering
\includegraphics[scale=0.5]{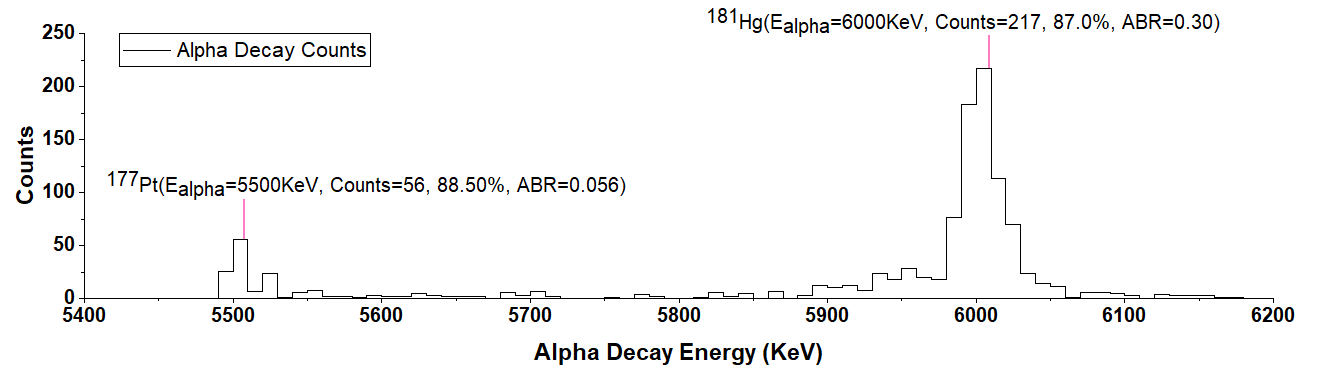}
\caption{$\alpha$-decay energy spectrum of $^{181}Hg$ and its decay products.}
\label{alpha-decay energy spectrum of 181Hg and its decay products.}
\end{figure}

In Fig.\ref{alpha-decay energy spectrum of 181Hg and its decay products.}, it is depicted that $^{181}Hg$($t_\frac{1}{2}=3.54s$) is peaked at 6000 KeV, with probability of decaying in this energy being 87\% and ABR$=$0.30. While its $\alpha$-decay product $^{177}Pt$($t_\frac{1}{2}=11s$) is peaked at 5500 KeV, with probability to decay in this energy being 88.5\% and ABR$=$0.056.

\subsubsection{PRODUCTION OF $^{182}Hg$}
\begin{figure}[h]
\centering
\includegraphics[scale=0.49]{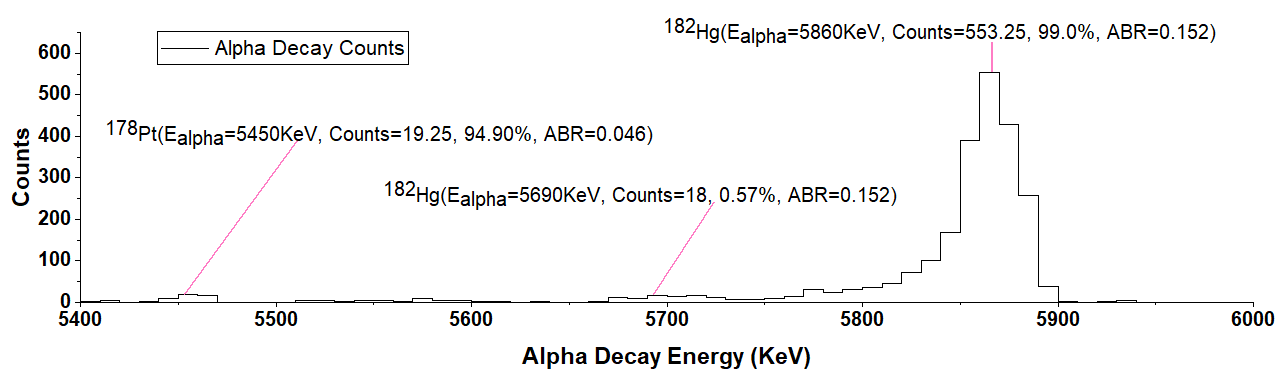}
\caption{$\alpha$-decay energy spectrum of $^{182}Hg$ and its decay products.}
\label{alpha-decay energy spectrum of 182Hg and its decay products.}
\end{figure}

In Fig.\ref{alpha-decay energy spectrum of 182Hg and its decay products.}, we can observe the base peak of $^{182}Hg$($t_\frac{1}{2}=10.83s$) at 5860 KeV. However, we can also see a peak of $^{182}Hg$ at 5690 KeV. This is due to the fact that a radioactive heavy nucleus can decay with multiple $\alpha$-decay energies with former being highly probable (99\%) and later being less probable (0.57\%). It is also observed that $\alpha$-decay product of $^{182}Hg$ is $^{178}Pt$($t_\frac{1}{2}=21.1s$) which is peaked at 5450 KeV.

\subsubsection{PRODUCTION OF $^{183}Hg$}
\begin{figure}[h]
\centering
\includegraphics[scale=0.49]{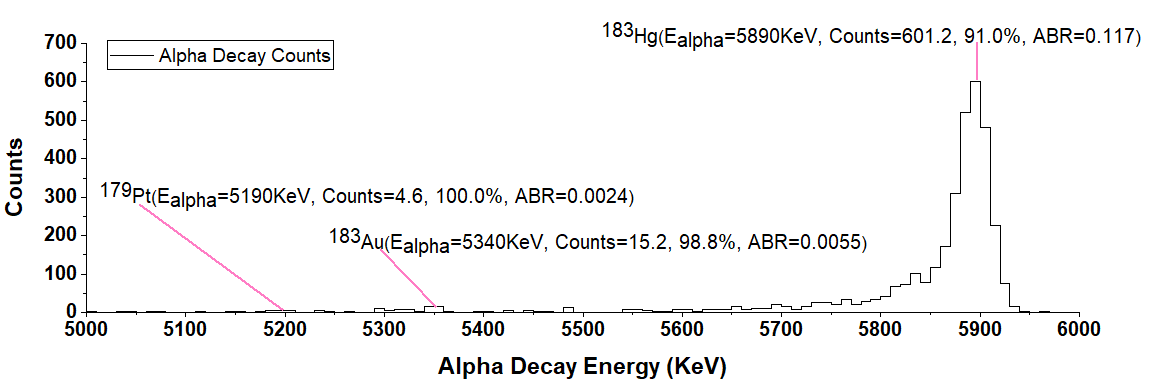}
\caption{$\alpha$-decay energy spectrum of $^{183}Hg$ and its decay products.}
\label{alpha-decay energy spectrum of 183Hg and its decay products.}
\end{figure}

In Fig.\ref{alpha-decay energy spectrum of 183Hg and its decay products.}, the two primary daughters of $^{183}Hg$, $^{183}Au$($t_\frac{1}{2}=42.8s$) and $^{179}Pt$($t_\frac{1}{2}=21.1s$) formed due to EC and $\alpha$-decay of parent nucleus are peaked at 5340 KeV, and 5190 KeV respectively. The parent nucleus $^{183}Hg$($t_\frac{1}{2}=9.4s$) is peaked at 5890 KeV.

\subsubsection{PRODUCTION OF $^{184}Hg$}
\begin{figure}[h]
\centering
\includegraphics[scale=0.49]{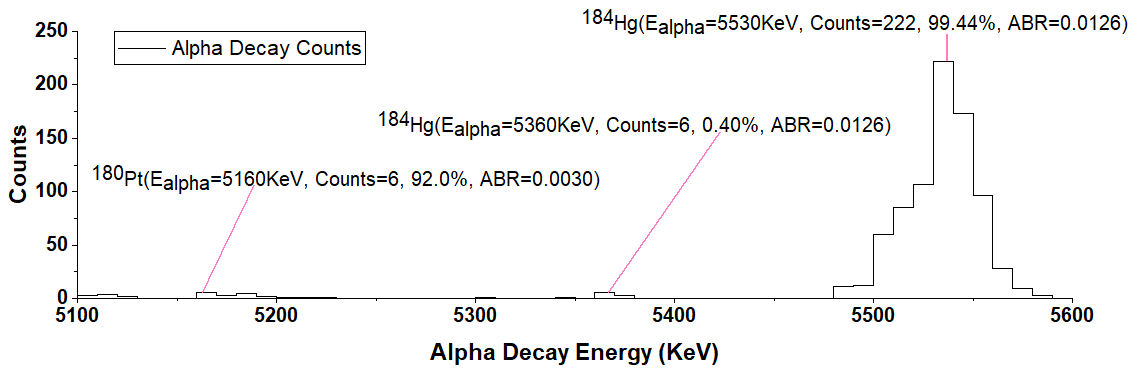}
\caption{$\alpha$-decay energy spectrum of $^{184}Hg$ and its decay products.}
\label{alpha-decay energy spectrum of 184Hg and its decay products.}
\end{figure}

In $\alpha$-decay energy spectrum of $^{184}Hg$ shown in Fig.\ref{alpha-decay energy spectrum of 184Hg and its decay products.}, the base peak corresponding to $^{184}Hg$($t_\frac{1}{2}=30.9s$) is obtained at 5530 KeV with probability of decaying with this energy being 99.44\%. While 0.4\% is the chance that $^{184}Hg$ decays with energy 5360 KeV. $^{180}Pt$($t_\frac{1}{2}=56s$) being the $\alpha$-decay product of $^{184}Hg$ is peaked at 5160 KeV.

\subsubsection{PRODUCTION OF $^{185}Hg$}
\begin{figure}[h]
\centering
\includegraphics[scale=0.52]{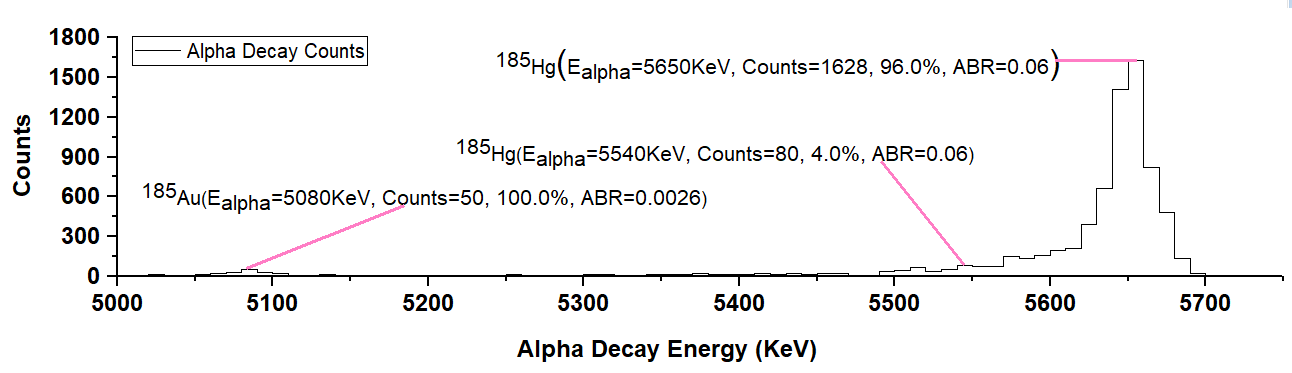}
\caption{$\alpha$-decay energy spectrum of $^{185}Hg$ and its decay products.}
\label{alpha-decay energy spectrum of 185Hg and its decay products.}
\end{figure}

In Fig.\ref{alpha-decay energy spectrum of 185Hg and its decay products.}, it is observed that $^{185}Hg$($t_\frac{1}{2}=49.1s$) is peaked at 5650 KeV and 5540 KeV with probability to decay with these energies being 96\% and 4\% respectively. The daughter nucleus formed due to EC in $^{185}Hg$ is $^{185}Au$($t_\frac{1}{2}=4.25 months$), which is peaked at 5080 KeV with ABR of 0.0026 in $\alpha$-decay energy spectrum. Also, the $\alpha$-decay product of $^{185}Hg$, which is $^{181}Pt$($t_\frac{1}{2}=52s$) was not observed in this spectrum because of its very low probability to undergo $\alpha$-decay. 
\clearpage

\subsubsection{HEATMAP OF Hg ISOTOPES}

\begin{figure}[h]
\centering
\includegraphics[scale=0.5]{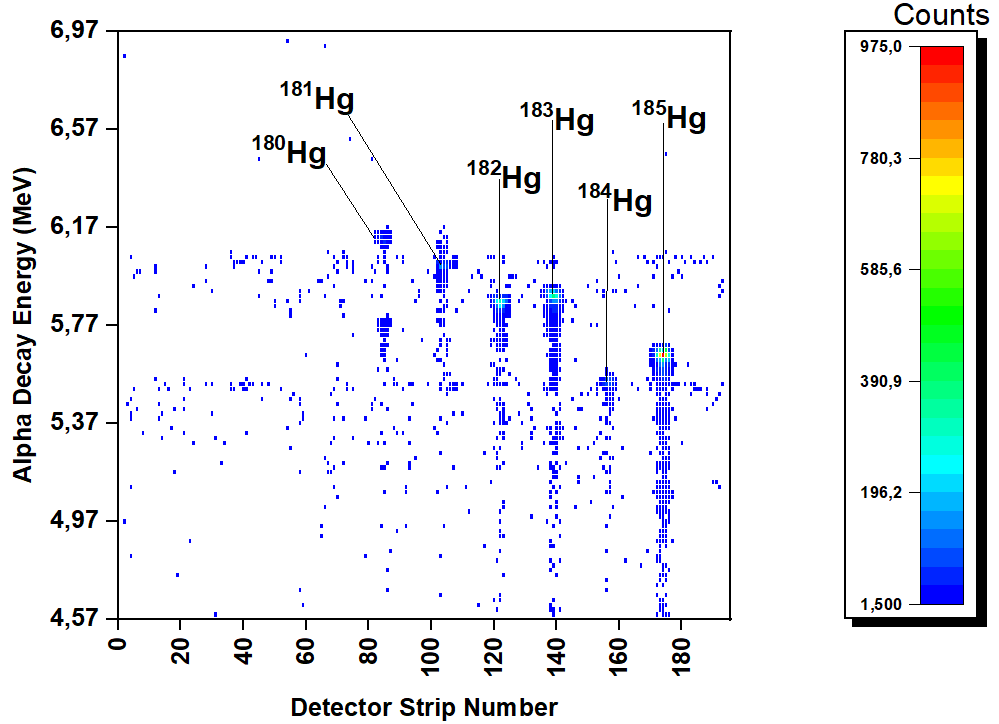}
\caption{Heatmap of Hg isotopes.}
\label{Heatmap of Hg isotopes.}
\end{figure}

Production of Hg isotopes in complete fusion reaction $^{148}Sm(^{40}Ar,xn)^{188-x}Hg$ is shown by a two-dimensional graph as shown in Fig.\ref{Heatmap of Hg isotopes.}. Through this energy-position graph, it's evident that different isotopes of Hg produced in aforementioned nuclear reaction are detected at unique strip numbers. However, only those isotopes of Hg are visible whose half-life is greater than average separation time (1.8±0.3 s) used by ISOL method for this reaction. Along the Y-axis we see $\alpha$-decay energy of each detected isotope and through the color associated with each isotope, we get the value of no. of counts of $\alpha$-particles detected by matching it with the color scale bar on right hand side.

The experimental values of $E_{alpha}$ is compared with its theoretical values obtained from \href{ https://drive.google.com/file/d/13T1d5-10oP43GgOamUdQ6NkXhaQ6NZ1J/view?usp=sharing}{table of nuclides} in Table \ref{Comparison between theoretical and experimental values of Ealpha of Hg isotopes}. It is observed that the \% change in their values is even less than 0.3\%. So, we conclude that the spectroscopic investigation performed for Hg isotopes in reaction $^{148}Sm(^{40}Ar,xn)^{188-x}Hg$ is almost accurate.
\begin{center}
\captionof{table}{Comparison between theoretical and experimental values of $E_{alpha}$(in KeV) of Hg isotopes produced in reaction $^{148}Sm(^{40}Ar,xn)^{188-x}Hg$.}
\begin{tabular}{c c c c}
\hline
Nucleus & Theo. $E_{alpha}$ & Exp. $E_{alpha}$ & $\Delta\%$ \\
\hline
$^{180}Hg$ & 6119 & 6120 & 0.016 \\
$^{181}Hg$ & 6006 & 6000 & 0.099 \\
$^{182}Hg$ & 5867 & 5860 & 0.119 \\
$^{183}Hg$ & 5904 & 5890 & 0.230 \\
$^{184}Hg$ & 5535 & 5530 & 0.090 \\
$^{185}Hg$ & 5653 & 5650 & 0.053 \\
\hline
\end{tabular}
\label{Comparison between theoretical and experimental values of Ealpha of Hg isotopes}
\end{center}

\section{SPECTROSCOPIC INVESTIGATION OF RADON ISOTOPES USING FULL-FUSION REACTION $^{166}Er(^{40}Ar,xn)^{206-x}Rn$}
A complete fusion reaction was performed between high energetic projectile particle ($^{40}Ar$) ejected from the window of U-400M cyclotron with an energy \~ 198 MeV and the target material $^{166}Er$ present in the form of rotating disc in the target box of MASHA facility. The products of the nuclear reaction were isotopes of Rn which were detected at focal plane (F2) of the position sensitive Si detector. Further, using the experimental data obtained from the detector and control system of MASHA, their $\alpha$-decay energy spectrum and energy-position graphs were plotted.

\subsubsection{PRODUCTION OF $^{201}Rn$}
\begin{figure}[h]
\centering
\includegraphics[scale=0.5]{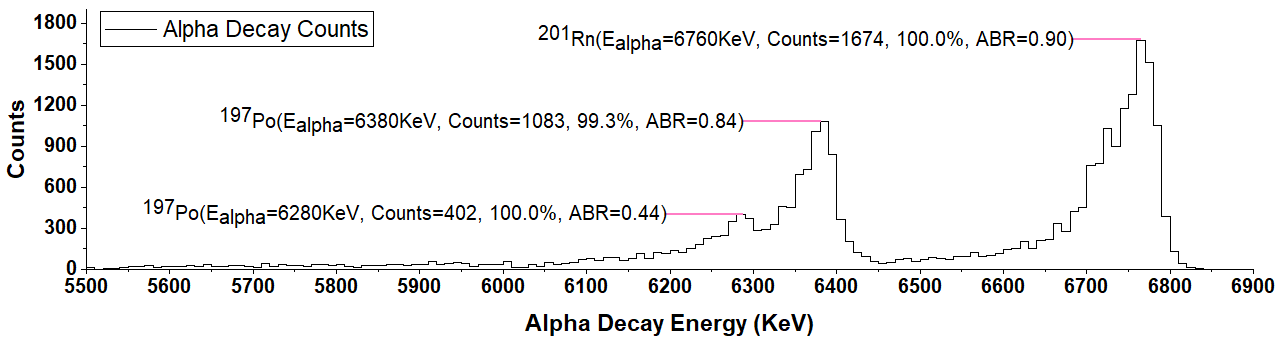}
\caption{$\alpha$-decay energy spectrum of $^{201}Rn$ and its decay products.}
\label{alpha-decay energy spectrum of 201Rn and its decay products.}
\end{figure}

In Fig.\ref{alpha-decay energy spectrum of 201Rn and its decay products.}, we see that the isotope $^{201}Rn$($t_\frac{1}{2}=3.8s$) is peaked at 6760 KeV. Its only daughter nucleus formed due to $\alpha$-decay, $^{197}Po$($t_\frac{1}{2}=53.6s$) with ABR$=$0.44 is peaked at 6280 KeV. However, we can see another $^{197}Po$($t_\frac{1}{2}=25.8s$) with ABR$=$0.84 peaked at 6380 KeV. The probability of occurrence of both $^{197}Po$($t_\frac{1}{2}=53.6s$) and $^{197}Po$($t_\frac{1}{2}=25.8s$) are very high with 100\% and 99.3\% respectively.

\subsubsection{PRODUCTION OF $^{202}Rn$}
\begin{figure}[h]
\centering
\includegraphics[scale=0.5]{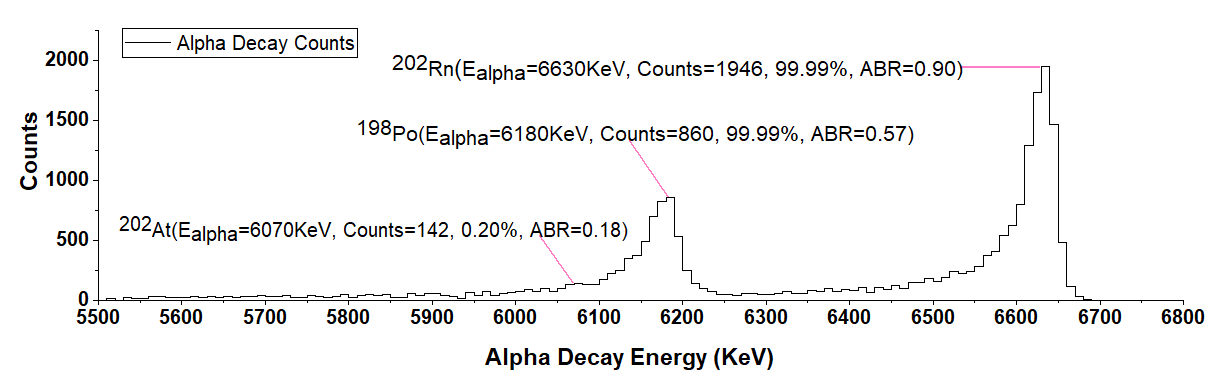}
\caption{$\alpha$-decay energy spectrum of $^{202}Rn$ and its decay products.}
\label{alpha-decay energy spectrum of 202Rn and its decay products.}
\end{figure}

As depicted in Fig.\ref{alpha-decay energy spectrum of 202Rn and its decay products.}, the two primary daughters of $^{202}Rn$, $^{202}At$($t_\frac{1}{2}=184s$) and $^{198}Po$($t_\frac{1}{2}=1.77 months$) formed due to EC and $\alpha$-decay of parent nucleus are peaked at 6070 KeV, and 6180 KeV respectively. The parent nucleus $^{202}Rn$($t_\frac{1}{2}=10.0s$) is peaked at 6630 KeV.

\subsubsection{PRODUCTION OF $^{203}Rn$}
\begin{figure}[h]
\centering
\includegraphics[scale=0.5]{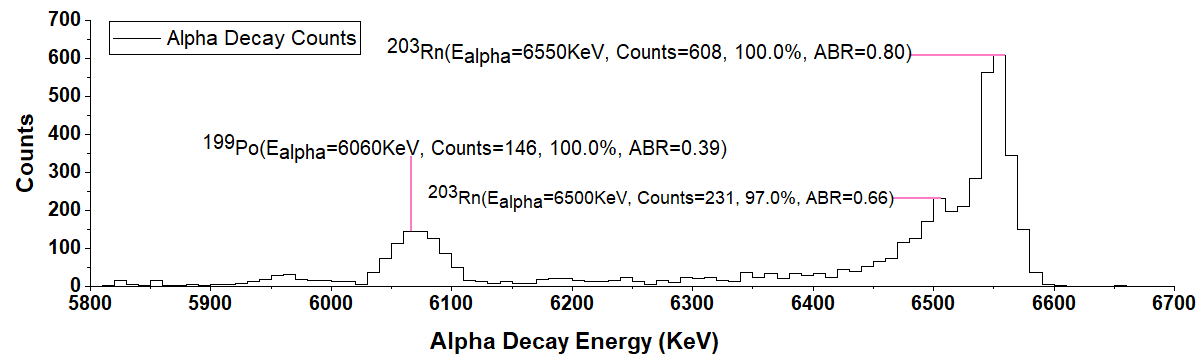}
\caption{$\alpha$-decay energy spectrum of $^{203}Rn$ and its decay products.}
\label{alpha-decay energy spectrum of 203Rn and its decay products.}
\end{figure}

During the investigation of $^{203}Rn$ isotope, it was observed that two different forms of $^{203}Rn$ with different $t_\frac{1}{2}$ and ABR were detected at same strip of position sensitive detector [see Fig.\ref{alpha-decay energy spectrum of 203Rn and its decay products.}]. $^{203}Rn$($t_\frac{1}{2}=28s$) with ABR$=$0.8 was peaked at 6550 KeV, while $^{203}Rn$($t_\frac{1}{2}=45s$) with ABR$=$0.66 was peaked at 6500 KeV. The daughter nucleus $^{199}Po$($t_\frac{1}{2}=4.17 months$) formed due to the $\alpha$-decay emission of $^{203}Rn$ is peaked at 6060 KeV in $\alpha$-decay energy spectrum.

\subsubsection{PRODUCTION OF $^{204}Rn$}
\begin{figure}[h]
\centering
\includegraphics[scale=0.5]{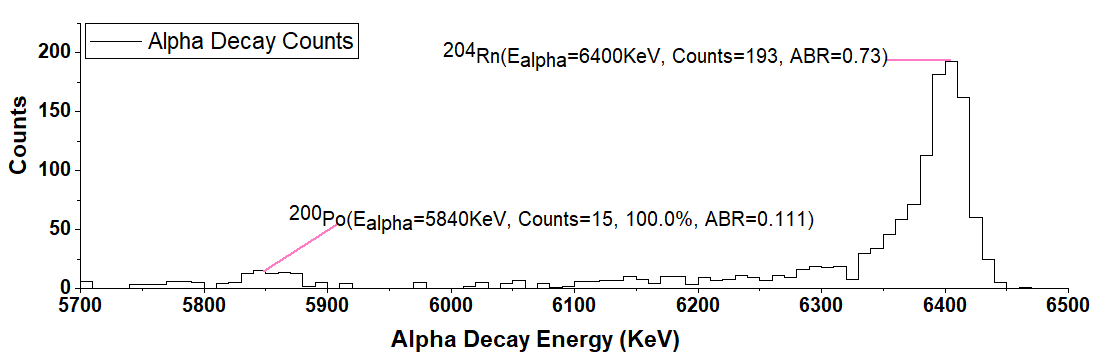}
\caption{$\alpha$-decay energy spectrum of $^{204}Rn$ and its decay products.}
\label{alpha-decay energy spectrum of 204Rn and its decay products.}
\end{figure}

From the $\alpha$-decay energy spectrum of $^{204}Rn$, it is noticed that the parent isotope $^{204}Rn$($t_\frac{1}{2}=1.24 months$) is peaked at 6400 KeV, while its $\alpha$-decay product $^{200}Po$($t_\frac{1}{2}=11.5 months$) is peaked at 5840 KeV as shown in Fig.\ref{alpha-decay energy spectrum of 204Rn and its decay products.}.   

\subsubsection{PRODUCTION OF $^{205}Rn$}
\begin{figure}[h]
\centering
\includegraphics[scale=0.5]{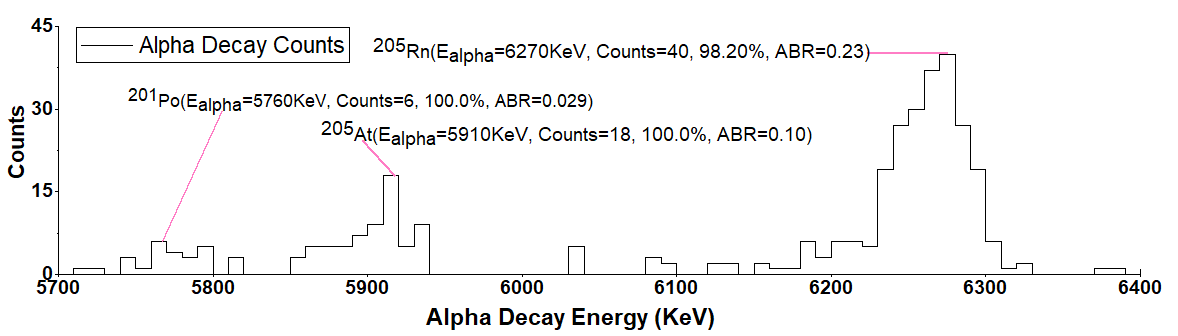}
\caption{$\alpha$-decay energy spectrum of $^{205}Rn$ and its decay products.}
\label{alpha-decay energy spectrum of 205Rn and its decay products.}
\end{figure}

From the Fig.\ref{alpha-decay energy spectrum of 205Rn and its decay products.}, it is evident that $^{205}Rn$($t_\frac{1}{2}=170s$) isotope undergoes $\alpha$-decay to release 6270 KeV of energy. The probability that it face $\alpha$-decay to produce $^{201}Po$($t_\frac{1}{2}=8.9 months$) with $E_{alpha}=5760 KeV$ is 0.23. However, this heavy radioactive nucleus also tries to achieve stability through electron capture, and for same the probability is 0.754 (as seen from table of nuclides). $^{205}At$($t_\frac{1}{2}=26.2 months$) which is the result of EC is peaked at 5910 KeV in $\alpha$-decay energy spectrum of $^{205}Rn$ isotope.

\subsubsection{HEATMAP OF Rn ISOTOPES ($^{166}Er(^{40}Ar,xn)^{206-x}Rn$)}

\begin{figure}[h]
\centering
\includegraphics[scale=0.5]{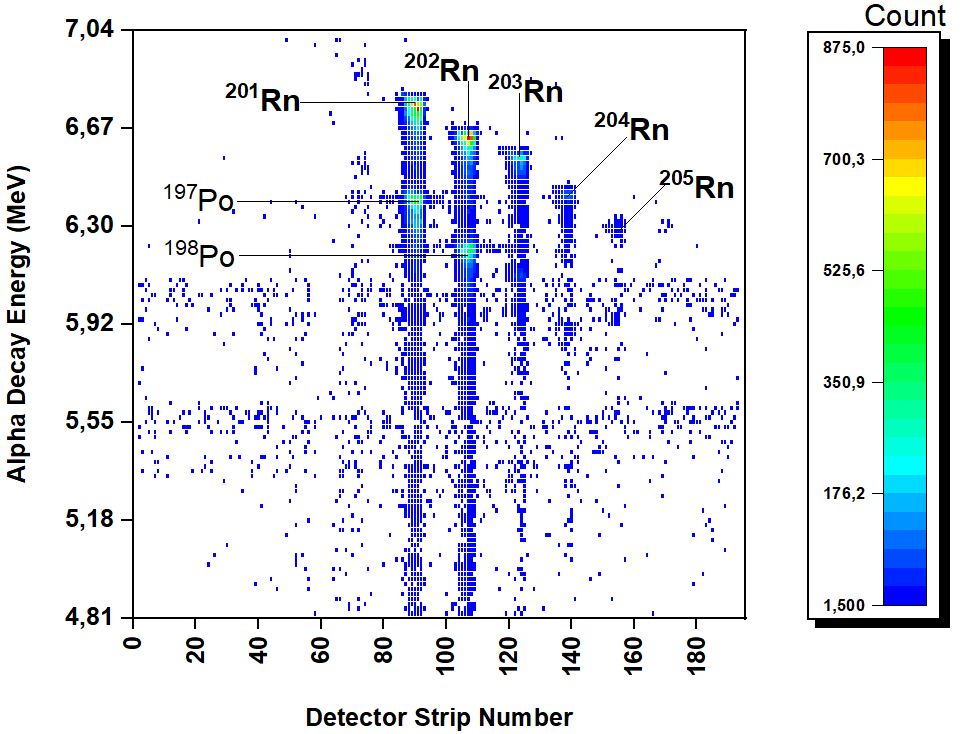}
\caption{Heatmap of Rn isotopes using complete-fusion reaction $^{166}Er(^{40}Ar,xn)^{206-x}Rn$.}
\label{Heatmap of Rn(201-205) isotopes.}
\end{figure}

The energy-position graph of Rn isotopes from $^{201}Rn$ to $^{205}Rn$ is shown in Fig.\ref{Heatmap of Rn(201-205) isotopes.}. These isotopes are the result of nuclear reaction $^{40}Ar$ + $^{166}Er$ $\rightarrow$ $^{206-x}Rn$ + $xn$ and are separated on the basis of their M/Q ratio in the magneto-optical system of mass-separator and finally detected at different strip numbers of detector. From this heatmap, we see $E_{alpha}$ decreases from $^{201}Rn$ to $^{205}Rn$. We also conclude that $^{201}Rn$ and $^{202}Rn$ isotopes are greatly produced in the reaction with high no. of counts, while $^{205}Rn$ is produced in relatively less number. The $\alpha$-decay daughter nuclei $^{197}Po$ and $^{198}Po$ from parent isotopes $^{201}Rn$ and $^{202}Rn$ respectively can also be seen at different strip numbers.

The comparison between theoretical and experimental values of $E_{alpha}$ for Rn isotopes produced in complete fusion reaction $^{166}Er(^{40}Ar,xn)^{206-x}Rn$ is shown in Table \ref{Comparison between theoretical and experimental values of Ealpha of Rn(201-205) isotopes}. It is noticed here that the \% change between these values is less than 0.3\%, so our analysis on this subject is nearly accurate.

\begin{center}
\captionof{table}{Comparison between theoretical and experimental values of $E_{alpha}$(in KeV) of Rn isotopes produced in reaction $^{166}Er(^{40}Ar,xn)^{206-x}Rn$.}
\begin{tabular}{c c c c}
\hline
Nucleus & Theo. $E_{alpha}$ & Exp. $E_{alpha}$ & $\Delta\%$ \\
\hline
$^{201}Rn$ & 6773 & 6760 & 0.192 \\
$^{202}Rn$ & 6639.5 & 6630 & 0.143 \\
$^{203}Rn(ABR=0.80)$ & 6549 & 6550 & 0.015 \\
$^{203}Rn(ABR=0.66)$ & 6499.3 & 6500 & 0.011 \\
$^{204}Rn$ & 6418.9 & 6400 & 0.294 \\
$^{205}Rn$ & 6262 & 6270 & 0.128 \\
\hline
\end{tabular}

\label{Comparison between theoretical and experimental values of Ealpha of Rn(201-205) isotopes}
\end{center}

\section{SPECTROSCOPIC INVESTIGATION OF RADON ISOTOPES USING MNTR $^{48}Ca$ + $^{242}Pu$}
Unlike complete fusion reactions discussed above, a MNTR can have any possible product nucleus. However, in the reaction of $^{48}Ca$ + $^{242}Pu$ under some fixed conditions, new neutron-rich Rn isotopes were produced near the neutron N=126 shell closure configuration, using MNTR. The isotopes produced were identified first, later their spectroscopic investigations were carried out. However, it was observed that only those Rn isotopes reached the detector and were identified which lived at least 35 ms while others decayed in their path.

\subsubsection{PRODUCTION OF $^{212}Rn$}
\begin{figure}[h]
\centering
\includegraphics[scale=0.5]{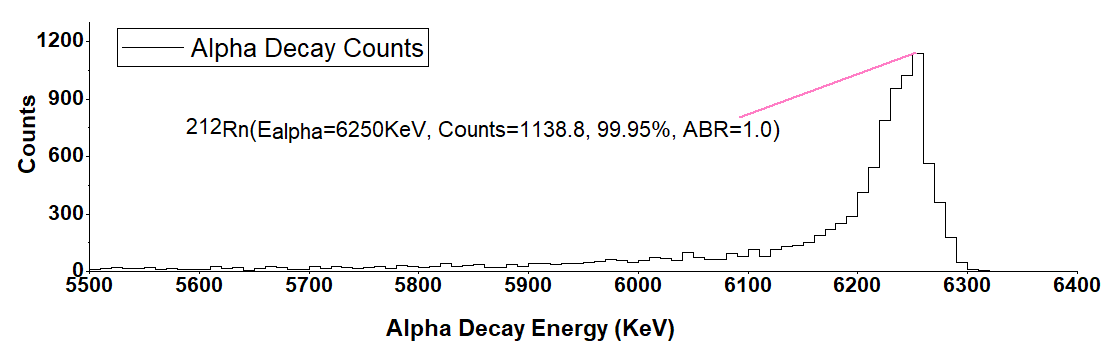}
\caption{$\alpha$-decay energy spectrum of $^{212}Rn$ and its decay products.}
\label{alpha-decay energy spectrum of 212Rn and its decay products.}
\end{figure}

From the spectrum of $^{212}Rn$($t_\frac{1}{2}=23.9 months$) isotope shown in Fig.\ref{alpha-decay energy spectrum of 212Rn and its decay products.}, it is evident that 99.95\% of $^{212}Rn$ isotopes undergoes $\alpha$-decay with energy 6250 KeV. Also, its ABR$=$1, which means it will only undergo $\alpha$-decay and no other forms of decay. $^{212}Rn$ isotope has N=126 shell configuration, which gives it a very high stability. Its daughter nucleus $^{208}Po$($t_\frac{1}{2}=2.9 years$)
was not observed in $\alpha$-decay energy spectrum.

\subsubsection{PRODUCTION OF $^{218}Rn$}

\begin{figure}[h]
\centering
\includegraphics[scale=0.5]{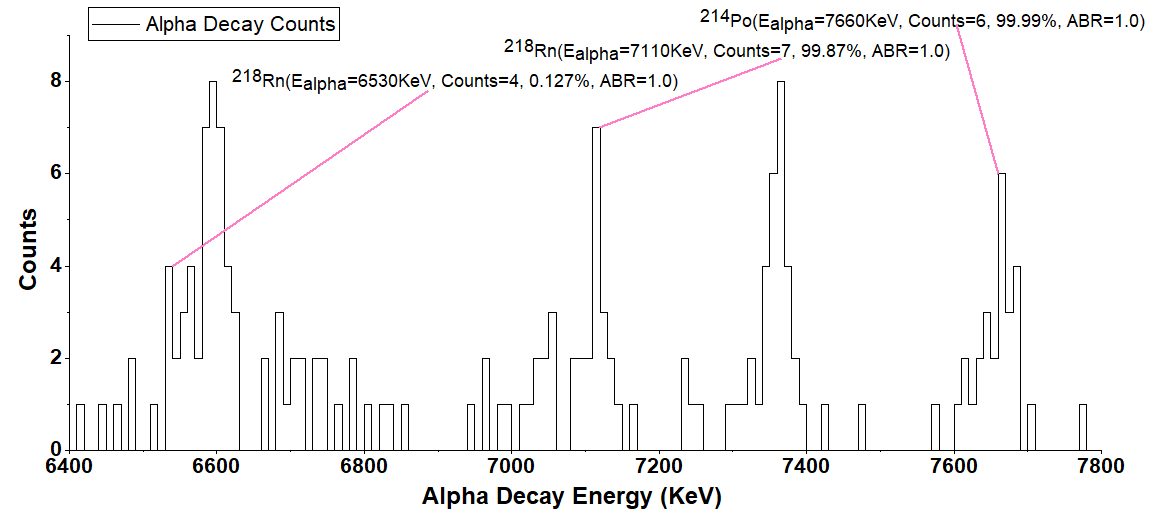}
\caption{$\alpha$-decay energy spectrum of $^{218}Rn$ and its decay products.}
\label{alpha-decay energy spectrum of 218Rn and its decay products.}
\end{figure}

The only daughter nucleus of $^{218}Rn$ due to its $\alpha$-decay is $^{214}Po$($t_\frac{1}{2}=164.3 \mu s$), which is peaked at 7660 KeV. The parent isotope $^{218}Rn$($t_\frac{1}{2}=35 ms$) is peaked at 7110 KeV and 6530 KeV with probability of decaying with these energies being 99.87\% and 0.127\% respectively. As discussed earlier, the reason behind why we see two different peaks of $^{218}Rn$ isotope in $\alpha$-decay energy spectrum is that the decay energy of alpha particle is not always fixed. It can have multiple energies, and there is a fixed probability for a certain energy to get released. Also, we see that the counts associated with $^{218}Rn$ isotope and its daughter nucleus is very low [see Fig.\ref{alpha-decay energy spectrum of 218Rn and its decay products.}] because of their low half-life.

\subsubsection{PRODUCTION OF $^{219}Rn$}
\begin{figure}[h]
\centering
\includegraphics[scale=0.5]{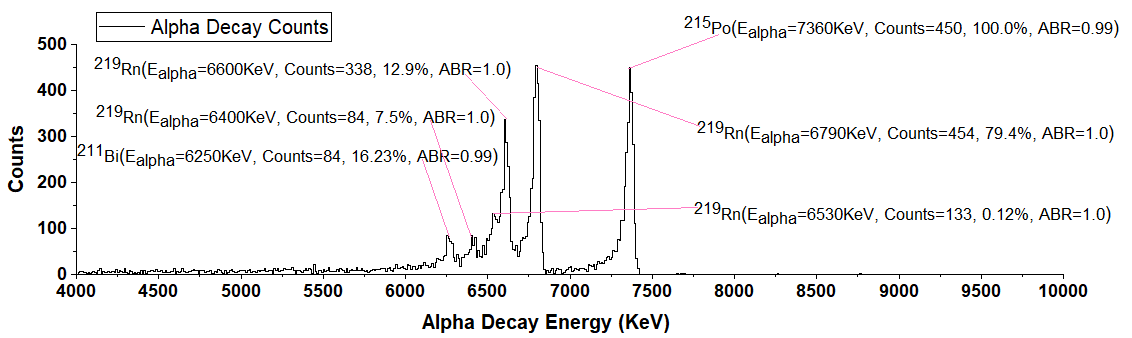}
\caption{$\alpha$-decay energy spectrum of $^{219}Rn$ and its decay products.}
\label{alpha-decay energy spectrum of 219Rn and its decay products.}
\end{figure}

From the $\alpha$-decay energy spectrum shown in Fig.\ref{alpha-decay energy spectrum of 219Rn and its decay products.}, we see that the parent isotope, $^{219}Rn$($t_\frac{1}{2}=3.96s$) is peaked at 6790 KeV, 6600 KeV, 6530 KeV, and 6400 KeV with probability to decay with these energies are 79.4\%, 12.9\%, 0.12\%, and 7.5\% respectively. All $\alpha$-decay energies have the same ABR$=$1. The $\alpha$-decay product of $^{219}Rn$ is $^{215}Po$($t_\frac{1}{2}=1.78 ms$), and is peaked at 7360 KeV. $^{211}Pb$($t_\frac{1}{2}=36.1 months$) which is the alpha-daughter of $^{215}Po$ could not be seen in this $\alpha$-decay energy spectrum because it only undergoes $\beta^-$ decay. However, its daughter nucleus, $^{211}Bi$($t_\frac{1}{2}=2.14 months$) is peaked at 6250 KeV, with ABR$=$0.99 and probability to decay with this energy is 16.23\%.

\subsubsection{HEATMAP OF Rn ISOTOPES ($^{48}Ca$ + $^{242}Pu$)}

\begin{figure}[h]
\centering
\includegraphics[scale=0.5]{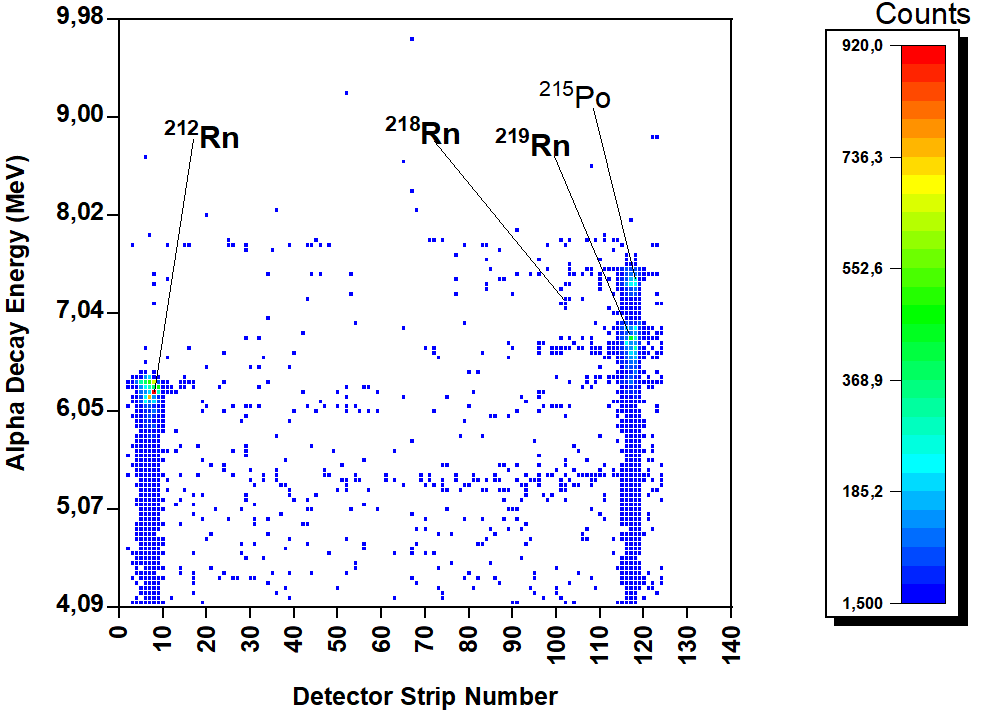}
\caption{Heatmap of Rn isotopes using MNTR $^{48}Ca$ + $^{242}Pu$.}
\label{Heatmap of Rn(212, 218, 219) isotopes.}
\end{figure}

The energy-position graph of Rn isotopes produced in MNTR $^{48}Ca$ + $^{242}Pu$ is shown in Fig.\ref{Heatmap of Rn(212, 218, 219) isotopes.}. $^{212}Rn$ and $^{219}Rn$ isotopes were largely produced in the reaction with high counts, while $^{218}Rn$ isotope with few counts is barely visible. This is because of low half-life of $^{218}Rn$($t_\frac{1}{2}=35 ms$), that major part of it decayed before reaching the detector. The $\alpha$-daughter of $^{219}Rn$, $^{215}Po$ was also detected at some strip number. A huge gap between $^{212}Rn$ and $^{218}Rn$ isotopes could be explained in terms of less half-life of Rn isotopes from A$=$213-217. Their half-life is even less than 35 ms which is more or less the average separation time used by ISOL method for this reaction. So, we conclude that only long-lived Rn isotopes were detected which lived at least 35 ms \cite{vedeneev2017current}, while others decayed before reaching the detector.

\begin{center}
\captionof{table}{Comparison between theoretical and experimental values of $E_{alpha}$(in KeV) of Rn isotopes produced in reaction $^{48}Ca$ + $^{242}Pu$.}
\begin{tabular}{c c c c}
\hline
Nucleus & Theo. $E_{alpha}$ & Exp. $E_{alpha}$ & $\Delta\%$ \\
\hline
$^{212}Rn$ & 6264 & 6250 & 0.223 \\
$^{218}Rn$ & 7129.2 & 7110 & 0.269 \\
$^{219}Rn$ & 6819.1 & 6790 & 0.427 \\
\hline
\end{tabular}
\label{Comparison between theoretical and experimental values of Ealpha of Rn(212, 218, 219) isotopes}
\end{center}

The theoretical and experimental values of $E_{alpha}$ for Rn isotopes produced via MNTR $^{48}Ca$ + $^{242}Pu$ is summarized in Table \ref{Comparison between theoretical and experimental values of Ealpha of Rn(212, 218, 219) isotopes}. $<0.3\%$ change in these values is observed for $^{212}Rn$ and $^{218}Rn$ isotopes, while for $^{219}Rn$, the \% change is $<0.5\%$. So, our analysis on this reaction is almost accurate.

\section{RESULTS AND CONCLUSIONS}
In this entire work, the production and spectroscopic investigation of Hg and Rn isotopes was performed using full fusion reactions $^{148}Sm(^{40}Ar,xn)^{188-x}Hg$, $^{166}Er(^{40}Ar,xn)^{206-x}Rn$  and multi-nucleon transfer reaction $^{48}Ca$ + $^{242}Pu$. The final product in all these reactions were isotopes of Hg and Rn. The experimental data obtained from the MASHA setup were analysed and 1D $\alpha$-decay energy spectrum graphs were plotted for those strips of detector which had detected any isotopic product of nuclear reaction. Further, this 1D histograms were used to plot a 2D energy-position graph, separately for Hg and Rn isotopes. The masses of super-heavy nuclei which were detected at different strips of Si based Position Sensitive Detector (PSD), have also been identified. Using 1D histograms and nuclide chart, the values of $E_{alpha}$, ABR, Counts, and the probability to decay with a specific amount of energy were calculated for all isotopic products of nuclear reactions studied in this paper.

\section{ACKNOWLEDGEMENTS}
The author expresses his gratitude to Mr. Viacheslav Vedeneev, Flerov Laboratory of Nuclear Reactions, Joint Institute for Nuclear Research for providing experimental data obtained from the MASHA setup. The author is also grateful to his parents for motivating him throughout his research work.

\section{STATEMENTS AND DECLARATIONS}
The author has no competing interests to declare that are relevant to the content of this article.

\bibliographystyle{elsarticle-num}
\bibliography{refer}
\end{document}